\documentclass[conference,a4paper]{IEEEtran}
\makeatletter
\def\ps@headings{%
\def\@oddhead{\mbox{}\scriptsize\rightmark \hfil \thepage}%
\def\@evenhead{\scriptsize\thepage \hfil \leftmark\mbox{}}%
\def\@oddfoot{}%
\def\@evenfoot{}}
\makeatother
\usepackage{multicol}
\usepackage{cite}
\usepackage[dvips]{graphicx}
\usepackage{epsfig}
\graphicspath{{eps/}}
\usepackage{subfigure}
\usepackage{float} 
\usepackage{amsmath}
\usepackage{array}
\usepackage{algorithm}
\usepackage{algorithmic}
\usepackage{mdwmath}
\usepackage{mdwtab}
\usepackage{url}
\newcounter{MYtempeqncnt}

\begin{document}
%\begin{multicols}{3}
\title{Adaptive Beamwidth Selection for Contention Based Access Periods in Millimeter Wave WLANs}
\author {\IEEEauthorblockN{Kishor Chandra\IEEEauthorrefmark{1}, R. Venkatesha Prasad\IEEEauthorrefmark{1}, I.G.M.M. Niemegeers\IEEEauthorrefmark{1}, Abdur R. Biswas\IEEEauthorrefmark{2}}
\IEEEauthorblockA{\IEEEauthorrefmark{1}EEMCS,Delft University of Technology, The Netherlands\\
%Mekelweg 4, 2628 CD, Delft, The Netherlands \\
Email: \{k.chandra, i.g.m.m.niemegeers\} @tudelft.nl, rvprasad@ieee.org.}
\IEEEauthorblockA{\IEEEauthorrefmark{2}Creat-Net, Italy\\
Email: abdur.rahim@create-net.org.}
}
%Telephone: (800) 555--1212, Fax: (888) 555--1212}
\maketitle
\begin{abstract}
%\boldmath
60\,GHz wireless local area networks (WLANs) standards (e.g., IEEE 802.11ad and IEEE 802.15.3c) employ hybrid MAC protocols consisting of contention based access using CSMA/CA as well as dedicated service periods using time division multiple access (TDMA). To provide the channel access in the contention part of the protocol, quasi omni (QO) antenna patterns are defined which span over the particular spatial directions and cover a limited area around access points. In this paper, we propose an algorithm to determine the beamwidth of each QO level. The proposed algorithm takes into account the spatial distribution of nodes to allocate the beamwidth of each QO level in an adaptive fashion in order to maximizes the channel utilization and satisfy the required link budget criterion. Since the proposed algorithm minimizes the collisions, it also minimizes the average time required to transmit total packets in a QO level. Proposed algorithm improves the average channel utilization up to 20-30$\%$ and reduces the time required to transmit total packets up to 40-50$\%$ for the given network parameters.
\end{abstract}
% Recently, several standards(e.g., IEEE 802.15.3c and IEEE 802.11ad) are proposed which define the PHY and MAC specifications.
% \fi
\IEEEpeerreviewmaketitle
\vspace{-2.0mm}
\section{Introduction}
The widely prevalent use of handheld devices such as smartphones and tablets coupled with users' demand for high quality video services and online gaming applications are forcing an immediate need for high speed (multi-Gbps), high quality, reliable and affordable communication technology. A large bandwidth at millimeter wave frequency band (30 to 300\,GHz) is available and is being investigated. In particular, 60\,GHz band has been of special interest for high speed (in the order of several Gbps), short range wireless communication.

The wave propagation in 60\,GHz is significantly different from 2.45\,GHz and 5\,GHz signals. Firstly 60\,GHz wireless propagation is subject to a very high transmission path loss. Considering a free space path loss (path loss exponent = 2), path loss at 60\,GHz is at least 20\,dB worse than that of 5\,GHz. High oxygen absorption ($10-15$\,dB/km) is another issue at 60 GHz frequency band, though prominent in outdoor environments at a distance of more than 100\,m~\cite{iee:smulders}. Another important characteristic is its limited ability to diffract around the obstacles making it unsuitable for non-line of sight  communication~\cite{iee:diffraction}.
%The received power $P_{r}(d)$ at a distance $d$ from an isotropic antenna is given as, 
%\begin{equation}
%P_{r}(d)=G_{t}G_{r}P_{t}(\frac{\lambda}{4\pi})^{2}(\frac{1}{d})^{\alpha}
%\end{equation}
%Where $G_{t}$ and $G_{r}$ are transmit and receive antenna gains, respectively, $P_{t}$ is transmitted power, $\lambda$ is the career wavelength and $\alpha$ is the path loss coefficient. For 60\,GHz frequency band $\lambda$ is much lower than that of 2.4\,GHz which results in very high free space path loss.

The high attenuation due to path loss and absorption can be mitigated by using directional antennas to overcome the high path loss and to provide high spatial multiplexing capability~\cite{iee:ref2}. Due to the smaller wavelength at 60\,GHz, compact directional antenna arrays can be easily implemented. To overcome frequent disruption by obstacles, intelligent beam switching~\cite{beamswitching_xeuli}, alternate path selection~\cite{diversity}, support of relay devices~\cite{relaybyZulkuf}  and radio over fiber based architecture~\cite{iee:ref7},\cite{iee:toward} are proposed.

The solutions proposed above for the physical layer issues have a strong impact on the design of the medium access control and higher layers. Several solutions are being standardized  to achieve multi-Gbps Wireless LAN (WLAN) at the 60 GHz band.IEEE 802.15.3c working group ~\cite{iee:IEEE802.15.3c} was formed to standardize MAC and PHY specifications at 60 GHz with data rates up to 5\,Gbps. Further, IEEE 802.11ad~\cite{IEEE802.11ad} aims to support data rates upto 7\,Gbps with a 2\,GHz frequency bandwidth at 60\,GHz. Along with the high data rate, it also provides compatibility with the popular IEEE 802.11 series (at 2.4\,GHz), making it the most preferable choice among available 60\,GHz standards.

\begin{figure}[!t]
\centering
\includegraphics[width=2.0in,height=1.7in]{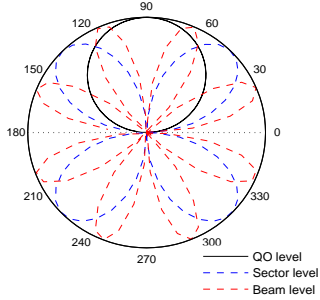}
\caption{Different antenna beamwidth patterns defined in IEEE 802.11ad.}
\label{antennapattern}
\vspace*{-5mm}
\end{figure}

 IEEE 802.15.3c~\cite{iee:IEEE802.15.3c} and IEEE 802.11ad~\cite{IEEE802.11ad} has proposed hybrid MAC protocols that consist of  carrier sense multiple access with collision avoidance (CSMA/CA) and time division multiple access (TDMA) based service periods allocation. To ensure the coverage, area around access point (AP) is divided into several level of beam widths with different granularity. For example, in IEEE 802.11ad, these levels are called the quasi omni, sector and beam levels. Fig.~\ref{antennapattern} depicts these different antenna beamwidth levels. Quasi omni (QO) patterns have the widest beam width followed by sector level having finer beamwidth, and beam levels have very narrow beamwidth. Usually, QO patterns are used during contention based access periods (CBAP), while fine beams (sector and beam levels) are used during TDMA based service periods. Slots in service periods are reserved by the devices beforehand, and any other device which may cause potential interference, is not allowed to access the medium. On the other hand, during CBAP, devices under a QO pattern simultaneously contend for the transmission opportunity. Thus, contention access period plays an important role in the protocols. As discussed above that area surrounding an AP is divided into three level of granularity, i.e., QO, sector and beam levels. According to IEEE 802.11ad/802.15.3c, CBAP allocation is done for each QO pattern in a time sharing basis. However, it is not defined as to how these QO levels are formed. Moreover it is assumed that all of the QO levels have equal and fixed beamwidths. 

Most of the work available in literature on performance analysis of 60 GHz MAC protocols have assumed the fixed/equal beamwidth QO levels. Pyo and Harada~\cite{harada}, have provided a detailed performance analysis of IEEE 802.15.3c and concluded that contention access periods have significant impact on overall system throughput. Further, Leonardo et al.~\cite{optbeam_leonardo} discussed the relation between link budget and sector level beamwidth, however, they focus on device association procedure. Chen et al.~\cite{11adcoperative} have incorporated cooperative relaying with IEEE 802.11ad MAC protocol. A relay selection mechanism is proposed to enhance the data rate, however, equal beamwidth QO levels are assumed in the analysis. In this paper, we propose an algorithm for formation of QO levels which takes into account the spatial distribution of nodes and the link budget to decide the beamwidth of access point in CBAP duration. As a result of this scheme, QO levels in different spatial directions can have different beamwidths. The scheme is based on the principle that MAC throughput of CSMA/CA protocol depends on the number of nodes competing for the channel time while actual data rate depends on how far the device is from the AP. In this article, we focus mainly on IEEE 802.11ad,  but the mechanism is general enough to be applicable to the IEEE 802.15.3c MAC and also for any other MAC protocol which employs directional antennas and CSMA/CA on time sharing basis in different spatial directions around AP. Thus, we provide a novel scheme for QO level beamwidth selection which will directly help in deciding the total number of QO levels required. Distribution of nodes as a basis for beamwidth selection helps in increasing the channel utilization in the contention phase of MAC protocol which, ultimately, will improve the overall throughput of WLAN system.
The main contributions of this article are:  
(i)~derivation of analytical expression for channel utilization in time shared CSMA/CA; (ii)~proposal for an adaptive beamwidth allocation mechanism for 802.11ad WLAN; (iii)~comparison of adaptive beamwidth allocation with fixed, equal beamwidth allocation in terms of channel utilization; (iv)~ derivation of total required time for contention based access period; and (v)~comparison of the average required time to complete the given number of requests using adaptive beamwidth allocation with fixed beamwidth allocation. 

In Section \ref{introlysis}, we provide a brief introduction of IEEE 802.11ad which is followed by analysis of time shared CSMA/CA protocol. Later, Section \ref{adpatbeam} describes the adaptive beamwidth allocation algorithm and calculation of estimated time required for contention based access periods. Further, in section \ref{discussion}, numerical results of beamwidth allocation algorithm are discussed. We conclude in Section \ref{remark}.

\section{MAC protocol description and channel utilization}\label{introlysis}
%In this section, a brief account of IEEE 802.11ad system model and MAC protocol specifications are presented. Then, we analyze the channel utilization by the MAC.
\subsection{System model and IEEE 802.11ad  MAC protocol}\label{intro}
\begin{figure}[]
\centering
\includegraphics[width=1.6in,height=!]{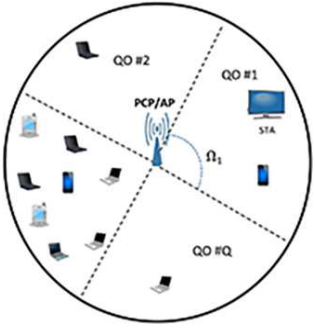}
 \caption{IEEE 802.11ad system model.}
\label{fig_system_model}
\end{figure}

IEEE 802.11ad defines a PBSS (Personal Basic Service Set) which is the operating area of network formed by 60\,GHz wireless stations (STAs), and is comparable to the IBSS (Independent Basic Service Set) at 2.4\,GHz and 5\,GHz. Unlike the IBSS, one of the STAs in a PBSS works as PCP/AP (PBSS control point/Access Point) to coordinate the channel access among STAs in the PBSS. STAs operating at 60\,GHz are called as DMG-STAs (Directional Multi Giga-bit STAs)\footnote{For simplicity, all the STAs are DMG-STAs and are referred as STAs. }. Fig.~\ref{fig_system_model} shows an IEEE 802.11ad PBSS where PCP is in the centre of the circle and STAs are distributed around the area covered by the PCP. Typical radius of a PBSS is about 10-20\,m. 

\begin{figure}[!t]
\centering
\includegraphics[width=2.8in,height=0.5in]{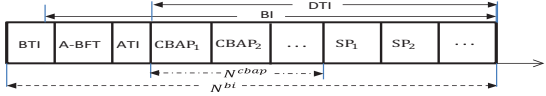}
 \caption{IEEE 802.11ad BI structure.}
 \label{fig_SF}
\end{figure}

Timing in IEEE 802.11ad is based on beacon intervals (BIs) set by PCP/AP. Duration within two beacon intervals is divided into different access periods having different medium access rules. Fig.~\ref{fig_SF} illustrates the different access periods within a beacon interval which comprised of : (i)~BTI (beacon transmission interval), during which beacons are transmitted; (ii)~A-BFT (association beamforming training), during which member STAs train their beams with PCP/AP; (iii)~ATI (announcement time interval), for exchange of management information between PCP/AP and the member STAs; and (iv)~DTI (data transfer interval), during which data transfer happens; and it consists of CBAPs (contention-based access periods) and SPs (service periods). CBAPs employ CSMA/CA for channel access by STAs, while SPs are reserved using service period request (SPR) command after PCP/AP polls an STA during the ATI period.

In this paper we focus on uplink channel access during the CBAP period. It is assumed that each STA associated with PCP/AP has trained its beam during A-BFT period using sector level sweep (SLS) procedure. Once SLS phase is completed, the best transmit beams  between PCP/AP and other STAs are assumed to be known. Thus, PCP/AP knows the location of the STAs. Channel contention during CBAP in different QO levels is allowed in a time sharing basis. STAs under each QO level contend for channel using CSMA/CA mechanism during allotted CBAP period for that QO level. In the next section, we evaluate the channel utilization during CBAP for an arbitrary number of STAs and derive an expression for the average channel utilization in a BI duration. For simplicity, we assume that BI only contains CBAPs without any SPs. 
\subsection{Throughput calculation of time sharing contention based access}\label{analysis}
CBAP uses CSMA/CA for medium access. The different QO levels of PCP/AP are switched in a round robin fashion. As shown in Fig.~\ref{fig_system_model}, only those STAs that are within the beamwidth of the current PCP/AP QO level will contend for the channel during the current CBAP duration. Since PCP/AP is aware of the STAs that would contend for the channel time in the current QO level, the problem of deafness, that is very prominent in directional antennas, is eliminated. All the STAs in a QO level are closer to each other due to smaller beamwidths. With request to send (RTS) - clear to send (CTS) mechanism being employed, it reduces the problem of hidden terminals.

Let beamwidth of PCP/AP for $k^{th}$ QO level  having $n_{k}$ STAs be denoted by $\Omega_{k}$, where, $1\le k \le Q$ and $Q$ is the maximum number of QO levels. Let $\tau_k$ be the transmission probability of each device in $k^{th}$ QO level, which is assumed to be constant over all the time slots. Let $p_k$ be the probability of collision experienced by a packet given that it is transmitted on the channel in $k^{th}$ QO level. $p_k$ is also known as the \textit{conditional collision probability} and is assumed to be constant and independent of the number of retransmission attempts in a QO level. For simplicity, henceforth, we will represent $\tau_k$ and $p_k$ by $\tau$ and $p$, respectively (though each QO level can have different values for $\tau$ and $p$). In saturation condition (i.e., each STA always has a packet to send), the relation of $\tau$ and $p$ is given by,
\begin{equation}\label{eq:collision}
p=1-\left(1-\tau\right)^{n_{k}-1}.
\end{equation}
\begin{figure}[!t]
\centering
\includegraphics[width=2.7in,height=2.2in]{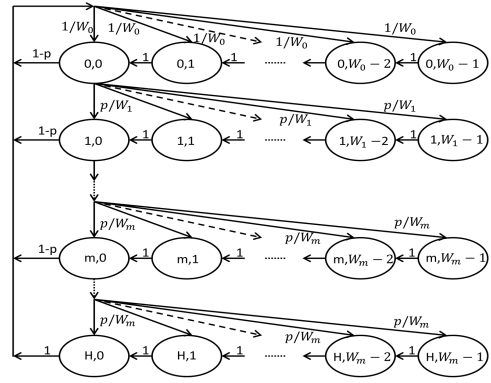}
 \caption{Markov chain model for packet transmission states.}
\label{fig_markov}
\end{figure}
% IEEE 802.15.3c CSMA/CA backoff window consists of discrete time backoff slots denoted as $pBackoffSlot$. 
%
 CSMA/CA employs random binary exponential back-off mechanism. If channel is idle, backoff counter decreases by one. When backoff counter of an STA reaches zero, it transmits the RTS frame. If RTS frame is successfully received by PCP/AP, after SIFS  duration, it responds with a DMG (direction multi gigabit) CTS  (DMG CTS) frame. After the successful reception of DMG CTS frame by the STA, communication link between the STA and PCP/AP is established. Since other STAs in the given QO level can hear DMG CTS frame, their backoff counters are suspended. After data transmission is completed, PCP/AP sends an acknowledgement (ACK) frame and the STA generates a fresh backoff counter for the next frame to be sent. If RTS frame collides, STA moves to next backoff stage until maximum backoff limit is reached and tries to transmit the frame until retry limit expires. 

Let maximum backoff stage be defined by $m$ and retry limit be defined by $H$. Let $W_{0}$ be the minimum window size, then window size at $i^{th}$ retransmission stage is  $W_{i}$=$2^{i}W_{0}$ until $W_{max}=2^mW_0$ is reached. After $m^{th}$ backoff stage, window size is kept $W_{max}$ till maximum retry limit $H$. After retry limit $H$, the packet is dropped. Thus, we have $W_i$=min($2^iW_0, W_{max}$), $(i\in[0,H])$. 

To analyze the CSMA/CA, Markov chain based analytical model has been widely used~\cite{iee:ref5}, \cite{iee:ref8}. We extend the  model of \cite{iee:ref8} by including time shared nature of IEEE 802.11ad protocol. We define a stochastic process $b(t)$  to represent backoff counter and $s(t)$ to represent the backoff stage. Since it is assumed that $\tau$ is constant over all the time slots for all the devices in a QO level and $p$ is constant and independent of the number of previously suffered collisions by the packet, a two dimensional Markov chain with states $(s(t),b(t))$, shown in Fig.~\ref{fig_markov}, is defined. After solving the Markov chain in Fig.~\ref{fig_markov}, the relation between $\tau$ and $p$ is given by~(\ref{eq23}),
\begin{equation}\label{eq23}
  \tau=\sum_{i=0}^{H}b_{i,0}=\frac{1-p^{m+1}}{1-p}b_{0,0},
   \end{equation}
   where, $b_{0,0}$ is given by~(\ref{eq22A}).
   \begin{figure*}[!t]
   % ensure that we have normalsize text
   \normalsize
   % Store the current equation number.
   \setcounter{MYtempeqncnt}{\value{equation}}
   \setcounter{equation}{2}
   \begin{equation}\label{eq22A}
    b_{0,0}=\frac{2(1-2p)(1-p)}{W_0(1-p)(1-(2p)^{m+1})+(1-2p)(1-p^{m+1}+(2^mW_0+1)(1-p^{H-m}))}
   \end{equation}
   % Restore the current equation number.
   \setcounter{equation}{\value{MYtempeqncnt}}
   % IEEE uses as a separator
   \hrulefill
   % The spacer can be tweaked to stop underfull vboxes.
   %\vspace*{4pt}
   %
   \end{figure*}
%\begin{equation}\label{eq:markov}
%\tau=\frac{2(1-2p)(1-p^{m+1})}{W(1-(2p)^{m+1})(1-p)+(1-2p)(1-p^{m+1})}.
%\end{equation}
Given this relation, (\ref{eq:collision}) and ~(\ref{eq23}) can be solved for $p$ and $\tau$ for the given network parameters. Once $p$ and $\tau$ are known, channel utilization is calculated as follows. \\
\begin{table}[!t]
%% increase table row spacing, adjust to taste
\renewcommand{\arraystretch}{1.0}
% if using array.sty, it might be a good idea to tweak the value of
% \extrarowheight as needed to properly center the text within the cells
\caption{CBAP Analysis Parameters.}
\label{table_CAP}
\centering
%% Some packages, such as MDW tools, offer better commands for making tables
%% than the plain LaTeX2e tabular which is used here.
\begin{tabular}{|c||c|}
\hline
RTS & 20 Octets \\
\hline
{DMG CTS} & 26 Octets \\
\hline
{ACK} & 14 Octets \\
\hline
SIFS & 2.5 $\mu$s\\
\hline
RIFS & 9 $\mu$s\\
\hline
DIFS & 13.5 $\mu$s\\
\hline
CCADetectTime & 4 $\mu$s\\
\hline
Minimum window size $W_0$ & 8 \\
\hline
Maximum backoff stage (m)  & 3 \\
\hline
Retry limit (H)  & 5 \\
\hline
data size  & 1024 octets \\
\hline
Transmit power $P_t$  & 10\,dBm \\
\hline 
Omni antenna gain G  & 1 (linear) \\
\hline
Fading loss $X_{\sigma}$  & 2\,dB \\
\hline
Receiver sensitivity (MC4)  & -64\,dBm \\
\hline
Receiver sensitivity (MCS0)  & -78\,dBm \\
\hline
Link margin  & 20\,dB \\
\hline
Path loss exponent $\alpha$  & 2 \\
\hline
\end{tabular}
\end{table}
As per CSMA/CA, a random time slot can either be idle or busy. Further, a busy slot either can results in a successful transmission or a collision. Thus, probability of a slot being idle, successful transmission or collision are given by $P_{idle}=\left(1-\tau\right)^{n_{k}}$, $P_{suc}=n_{k}\tau(1-\tau)^{n_{k}-1},$ and $P_{col}=1-n_{k}\tau(1-\tau)^{n_{k}-1}-\left(1-\tau\right)^{n_{k}}$, respectively.

Let $U_{k}$ be the channel utilization in $k^{th}$ QO level, which is defined as the fraction of time that the channel is used to transmit payload successfully. Let $T_{idle}$ be the duration of an idle time slot, $T_{suc}$ is the duration of a successful transmission and $T_{col}$ is the duration of a failed transmission, then $U_{k}$ is calculated as,
\setcounter{equation}{3}
\begin{align}\label{total1}
U_{k}&=\frac{P_{succ}E[Payload]}{P_{idle}T_{idle} + P_{suc}T_{suc} + P_{col}T_{col}},
\end{align}
where $E[Payload]$ is the average duration of a payload packet and $T_{idle}= SIFS + CCADetectTime $.
Here, $T_{suc} =T_{rts} + 2SIFS + T_{cts}+DIFS+T_{DATA}+T_{ACK}$ and  $T_{col} = T_{rts} + SIFS + DIFS+T_{out}$, where, $T_{rts}$ is the duration of RTS frame, $T_{cts}$ is the duration of DMG CTS frame,  $DIFS$ is the Back off inter-frame space and $T_{out}$ is the acknowledgement or response timeout. Assuming that all the QO patterns are allotted CBAPs in one BI, overall channel utilization during BI (assuming there are no SPs) can be given by,
\begin{equation}\label{total}
U=\frac{1}{Q}\sum_{k=1}^{k=Q}U_k.
\end{equation}
\begin{figure}[!t]
\centering
\includegraphics[width=2.0in,height=1.6in]{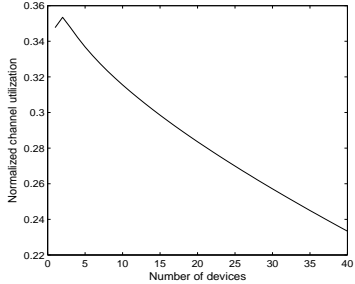}
 \caption{CBAP channel utilization in a QO level.}
\label{fig:sim}
\end{figure}
Fig.~\ref{fig:sim} shows the channel utilization with varying number of STAs for one QO level. All parameters used in the numerical evaluation are listed in Table~\ref{table_CAP}. IEEE 802.11ad Control PHY is used to transmit RTS, DMG CTS and ACK frames, while MCS4 (modulation and coding scheme 4) is used for data frame transmission. Size of a data frame is 1024\,Bytes. It can be easily seen that channel utilization decreases rapidly with increase in the number of STAs. In IEEE 802.11b/g WLAN, omnidirectional antennas are used which enables simultaneous medium access in an IBSS for all the member STAs. 
%To improve the channel utilization, several solutions are proposed such as controlling minimum and maximum window size and retransmission limits to reduce the collision probability which require that all the STAs adapt these parameters changes.
 However, IEEE 802.11ad employing directional antennas provides one extra degree of freedom (i.e. beamwidth) and can control the number of simultaneously contending STAs by restricting the channel access in some spatial directions while allowing in others. This is the motivation behind adaptive beamwidth allocation and a novel algorithm is proposedin the next section.
\section{Adaptive beamwidth and timing allocation}\label{adpatbeam}
IEEE 802.11ad/802.15.3c does not provide any mechanism for selecting the beamwidth of individual QO levels. It only defines them as QO levels~\cite{iee:beamcodebook}. Expression (\ref{total1}) for average channel utilization  during CBAB indicates the fraction of total time channel is being used for data transmission, which largely depends on the conditional collision probability $p$. Consequently, using the equal beamwidth QO levels can lead to very high collision probability in the densely populated regions or under utilization of channel in the QO levels having less number of STAs. To address this problem, we propose an algorithm for appropriate beamwidth selection for each QO level which tries to maximize CSMA/CA channel utilization.\\ 
\subsection{Beamwidth selection procedure}\label{algo}
We neglect beam switching time from one sector to another since switching antenna beam from one direction to another takes very little time compared to the time wasted in a collision. ~\cite{iee:MiuraOKU10} reports that an eight sector antenna array takes less than 100~$ns$ to switch from one sector to another sector. Also,~\cite{iee:IEEE802.15.3c} suggests an interval of 0.5 $\mu$s for beam switching (Beam switching interframe space time) which is much less than the time spend in a collision. 
.. 
Beamwidth $\Omega_{k}$ of $k^{th}$ QO level can vary from $\Omega_{min}$\footnote{ $\Omega_{min}$ is the minimum beamwidth.} to $\Omega_{max}$. The value of $\Omega_{min}$ is decided by the capability of antenna array to narrow down its beamwidth to the least possible value (highest possible beam resolution) and $\Omega_{max}$ is limited by the intended maximum distance the PCP/AP has to cover. If we consider a perfect conical antenna model, antenna gain $G(\Omega)$ for a beamwidth of $\Omega$ can be given by, $G(\Omega)=\frac{2\pi}{\Omega}$, where, $G$ is the gain of omnidirectional antenna. Thus, antenna beamwidth will decide the maximum distance it could cover for a particular MCS and transmitted power. Hence value of $\Omega_{max}$ depends on the data rate and intended distance to be covered by the transmitter.
  Link budget at the receiver which is distance $d$ apart from the transmitter can be given by,
  \begin{align}
  P_{r}(d)=&P_{t}+G(\Omega_t)+G(\Omega_r)-PL_0-10\alpha\log_{10}(d)\nonumber\\ &-X_{\sigma}-LM
   \end{align}
  Where, $P_t$ is transmitted power, $G(\Omega_t)$ and $G(\Omega_r)$ are transmit and receive antenna gains, respectively (which corresponds to the beamwidth ${\Omega_t}$ and ${\Omega_r}$ of transmit and receive antennas, respectively). $PL_{0} = 10\alpha\log_{10}(\frac{4\pi}{\lambda})$ is the reference path loss at a distance of 1\,m. $LM$ is the link margin, which is around $20$\,dB, used to compensate for localization errors~\cite{beaconImplement}. $X_{\sigma}$ is the fading loss and $\alpha$ is the path loss exponent.
  If $RS$ is the receiver sensitivity for a given modulation and coding scheme (MCS), then for satisfactory reception of signal, $P_r \geq RS$.  
  Value of $RS$ is given by the 802.11ad standard  for each MCS with a particular noise figure and packet error rate.\\
\begin{figure}[!t]
\centering
\includegraphics[width=2.8in,height=2.1in]{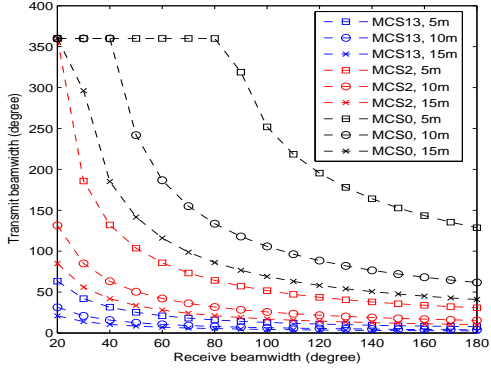}
 \caption{Required Rx/Tx beamwidth for different MCS and distances.}
\label{fig_rxtx}
\end{figure}
Fig.~\ref{fig_rxtx} shows the relation between required transmit antenna beamwidth for a given receiver beamwidth, MCS and intended coverage range. Curves are plotted for three modulation schemes for distances of 5\,m, 10\,m and 15\,m. It can be observed that low data rate MCS0 can operate at relatively wider beams as compared to higher order modulation schemes MC2 and MC13. In this way, for given STAs beamwidths, data rate and intended transmission distance, maximum allowed beamwidth $\Omega_{max}$ can be selected.
 To determine the beamwidth of QO levels, we start with the minimum antenna beamwidth $\Omega_{min}$ for the first sector and keep increasing the beamwidth by $\Delta\Omega$ until the throughput reaches its maximum value. With increasing beamwidth, first CSMA/CA throughput increases because initially channel utilization increases due to the increase in number of devices, but, when we keep increasing the beamwidth, after a certain beamwidth, CSMA/CA throughput starts decreasing due to the increase in collisions. The detailed algorithm is described in Algorithm~\ref{ALG}.
\begin{itemize}
\item {\textbf{Step 0)} \textit{Initialization (see line 1)}:

 During the association process (A-BFT duration), STAs discover the PCP/AP and train their beams with PCP/AP using SLS procedure. During this process, PCP/AP collects the angle information (i.e., angle w.r.t the position of PCP/AP) of all the STAs which are associated with the PCP/AP. Let $A$ represent the set of angular information $\beta_j (j=1,2 \dots, n )$ of all the $n$ STAs associated with the PCP/AP.}
 
\item {\textbf{Step 1)} \textit{evaluation for the minimum beamwidth $\Omega_{min}$ (see line 3)}:

 Starting with a beamwidth $\Omega_{min}$, PCP/AP calculates the number of devices in this beam area and CSMA/CA throughput. Rename this beamwidth as $\Omega_{p}$ (i.e., past value) and corresponding throughput as $U_p$.}

\item {\textbf{Step 2)} \textit{increment by the differential beamwidth $\Delta\Omega$ (see line 4)}:

 Take $\Omega_{p}$ and $U_p$ as reference values. Then, increase the beamwidth by $\Delta\Omega$, new beamwidth is denoted as $\Omega_n$ (since, $\Omega_n$ = $\Omega_p$ + $\Delta\Omega$, more devices are likely to be included in the increased beam area). Calculate the CSMA/CA throughput denoted as $U_n$.}

\item{\textbf{Step 3)} \textit{determine the appropriate QO beamwidth (see line 5 - 9)}:

  Compare $U_n$ and $\Omega_n$ with $U_p$ and $\Omega_{max}$, respectively. if $U_n \geq U_p$ and $\Omega_n \leq \Omega_{max}$, set $U_p = U_n, \Omega_p = \Omega_n$ and go to Step 3. Otherwise, select $\Omega_k =\Omega_p$ as the optimum beamwidth for sector $k$ (initially, k=1).} 

\item{\textbf{Step 4)} \textit{iteration to include all the STAs (see line 2 and 10)}:

 After deciding the beamwidth for $k^{th}$ sector, go to Step 2. Take the end point of previous sector as the starting point for the next sector. Repeat the same procedure to decide the beamwidth for ${k+1}^{th}$ sector. Do the same procedure until all the devices are included or the complete area around the PCP/AP is traversed.} 
\end{itemize}
\begin{algorithm}
\caption{Adaptive QO beamwidth selection}\label{ALG}
\begin{algorithmic}[1]
\STATE {\textbf{initialize} $\textit{A}:=\{\beta_{i}|1\leq i \leq n\}$, $\Omega_{min}$, $\Delta\Omega$ and $\Omega_{max} $};
\WHILE{$A\neq \emptyset$}
\STATE {$\Omega_p$$\leftarrow$ $\Omega_{min}$ and calculate $U_p$};
\STATE{$\Omega_n\leftarrow\Omega_p+\Delta\Omega$ and calculate $U_n$};
\IF{($U_n \geq U_p$ and $\Omega_n\le \Omega_{max}$)}
\STATE {$\Omega_p \leftarrow \Omega_{n}$, $U_p \leftarrow U_n$ and go to step 4};
\ELSE
\STATE $\Omega_k \leftarrow \Omega_p$;
\ENDIF
\STATE {go to step 2 and repeat the procedure for $(k+1)^{th}$ sector};
\ENDWHILE
\RETURN {$\Omega,n_k:=\{\Omega_k| 1\leq k \leq TotalSector$\}};
\end{algorithmic}
\end{algorithm}
%
%
%\vspace{-2.0mm} 
%
% 
\subsection{Computation of required CBAP duration for individual sectors} \label{cbapduration}
%According to IEEE 802.15.3c, CAP duration is usually 10\% of the overall SF duration. This is usually not the case while using RoF based architecture for indoor network because of extra propagation delay introduced by fiber. During CAP, each request and corresponding acknowledgement and/or response frames undergo the extra delay due to time elapsed in transmission through fiber which significantly widens the required CAP duration. On the one hand, if CAP duration is too long, it would reduce the CTAP duration as SF has a fixed maximum duration.  On the other hand, if CAP duration is too short, all the requests could not be successfully handled during CAP and there is a possibility that a device which wanted to release a CTA will leave that CTA unused. Thus the channel time is wasted. 

 Let $n_{id}$  represent the sum of average number of idle slots in each backoff stage and $n_{b}$ represent the sum of average number of busy slots in each backoff stage. Let $T_{id}$ and $T_{b}$ represent the idle time slot duration and average busy slot duration respectively. Further, $n_{b}=n_{col}+n_{suc}$ ($n_{col}$ and $n_{suc}$ are collision and successful slot counts respectively). Total CAP duration $T_{CBAP}$ can be written as,
\begin{equation}\label{eq:CAPmin}
%T_{CAP}=n_{id}T_{id} + n_{col}T_{col}+n_{suc}T_{suc}.
T_{CBAP}=n_{id}T_{id} + n_{b}T_{b}.
\end{equation}
Let there are total $N$ request to be completed. In order to ensure all the requests (on average) to be successfully transmitted, the sufficient and necessary condition is $n_{suc}\ge N$.

If $H$ is the maximum packet retry limit, the average backoff slots before a device successfully transmits a frame can be expressed as,
\begin{equation}\label{eq:sumavgdelay1}
n_{id}=\sum\limits_{0}^{H}P_{suc}(i)E[B_i], 
\end{equation}
where $E[B_i]$ is the average backoff slots accumulated by a device in backoff stage $i$. Assuming that backoff selection mechanism is uniformly distributed over the backoff window $[0, W_{i}]$, $E[B_i]$ can be expressed as,
\begin{align}\label{eq:avgbckoff}
E[B_i]=\frac{1}{2}
\begin{cases}
\begin{array}{ll}
\sum\limits_{0}^{i}2^{k}W_{0}, & 0\le i \le m,  \\
\sum\limits_{0}^{m}2^{k}W_{0} + (i-m)W_{max}, & m\le i \le H
\end{array}
\end{cases}
\end{align}
$P_{suc}(i)$ is the probability that a device transmits successfully in stage $i$ is given by $P_{suc}(i)=p^{i}\times (1-p),  i< H$. Because of finite number of retransmission attempts, probability that a packet reaches $H^{th}$ (maximum allowed) backoff stage is $p^{H}$. At this stage, it does not matter whether a packet is successfully transmitted or dropped after maximum retry limit is reached, because the packet will be removed from the system.
Therefore, now substituting the value of $E[B_i]$ from (\ref{eq:avgbckoff}) and  $P_{suc}$ in (\ref{eq:sumavgdelay1}), $n_{id}$ can be obtained.\\
%Therefore, now from (\ref{eq:sumavgdelay1}) and (\ref{eq:avgbckoff}),
%\begin{align}\label{eq:delaysucc}
%n_{i}=E[B]&=\sum\limits_{i=0}^{i=m-1}\left[p^{i}\times (1-p)\frac{1}{2}\sum\limits_{k=0}^{k=i}(2{^k}W_{0})\right]\nonumber\\
%&+p^{m}\frac{1}{2}\sum\limits_{k=0}^{k=m}(2{^k}W_{0}).
%\end{align}
%%
\begin{figure}
\centering
%\mbox{
\subfigure[CBAP channel utilization.]{ 
%\centering
\includegraphics[width=1.6in,height=1.4in]{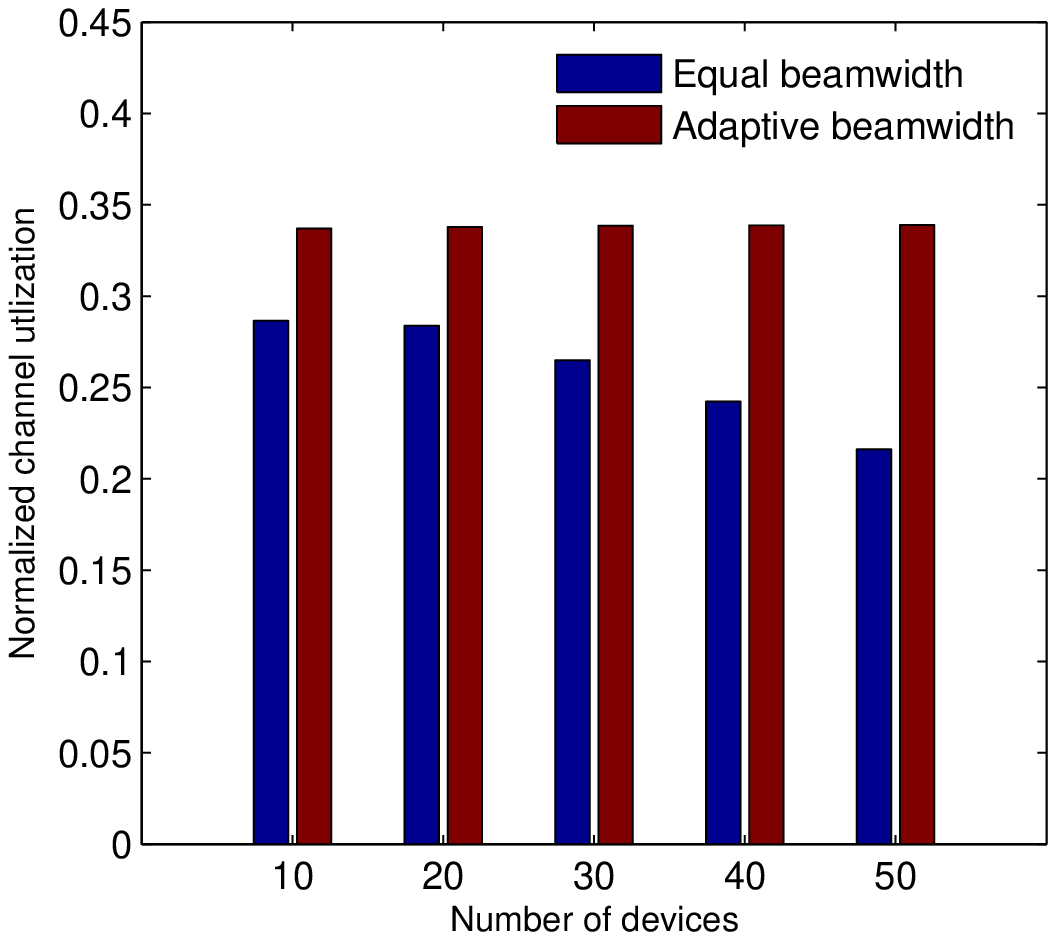}
\label{fig:CAPthroughputAdapt}
}
%\quad
%add desired spacing between images, e. g. ~, \quad, \qquad etc.
\subfigure[Minimum required CBAP duration.]{
%\centering
\includegraphics[width=1.6in,height=1.4in]{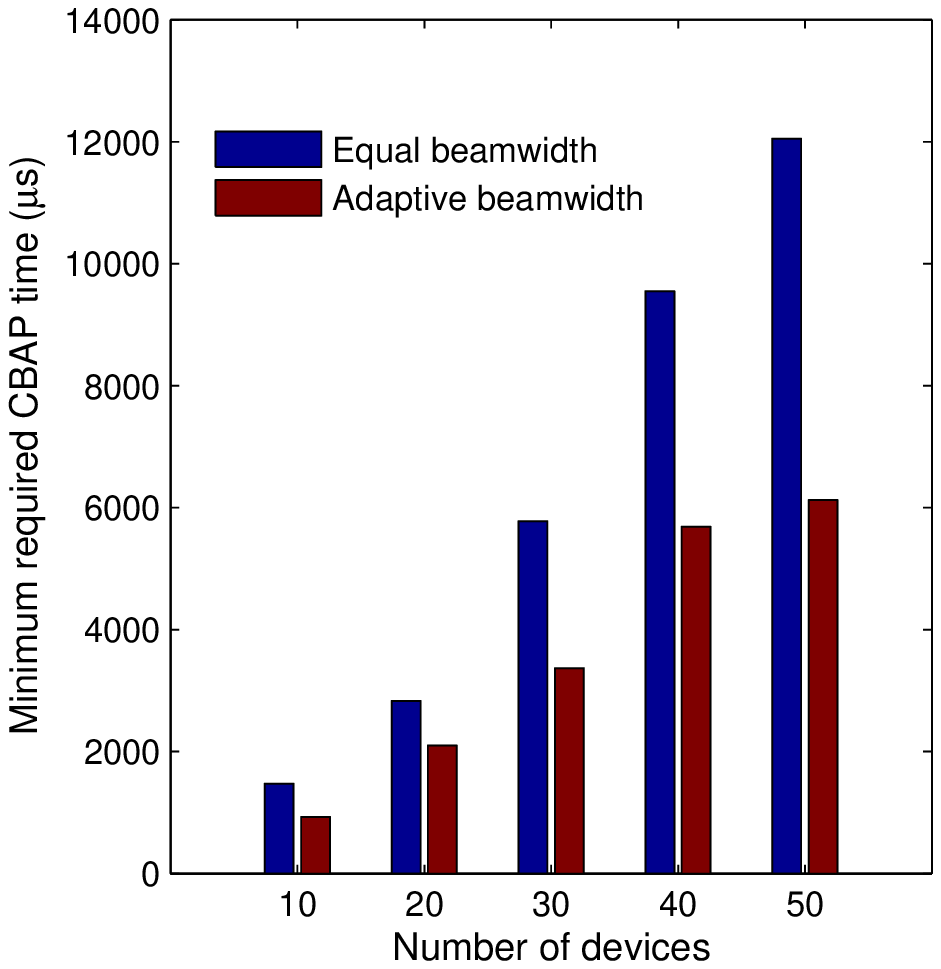}
\label{fig:TCAPminAdapt}
}
%}
\caption{CBAP performance comparison for adaptive and fixed beamwidth sectors.}
\label{fig:CAPcompare}
\end{figure}
Busy slots have two possible states, i.e., either collision or successful transmission. therefore, busy slots can be modelled as a Bernoulli random variable with $p_{suc/busy}$ and $p_{col/busy}$ as probability of a successful transmission and collision in a given busy time slot, respectively. Expressions for $p_{suc/busy}$ and $p_{col/busy}$ can be derived as,
\begin{equation}\label{eq:csr1}
p_{suc/busy}=\frac{P_{suc}}{1-P_{idle}}=\frac{n\tau(1-\tau)^{n-1}}{1-\left(1-\tau\right)^{n}}.
\end{equation}
And,
\begin{equation}\label{eq:csr2}
p_{col/busy}=\frac{P_{col}}{1-P_{idle}}=\frac{1-n\tau(1-\tau)^{n-1}-\left(1-\tau\right)^{n}}{1-\left(1-\tau\right)^{n}}.
\end{equation}
From the definition of Bernoulli trials, the expected number of successful slots can be given as, $n_{suc}=n_{b}p_{suc/busy}$. In order to minimize the CBAP duration we have to calculate the minimum required number of busy slots (denoted as $n_b[min]$), which would ensure that average number of successful slots become equal to the average number of CBAP requests $N$ in a given CBAP duration. Hence, $n_b[min]$ is equal to,
\begin{equation}\label{eq:csr6}
n_b[min]=\frac{N}{p_{suc/busy}}.
\end{equation}
On substituting value of $n_b$ from (\ref{eq:csr6}) and $n_{id}$ from (\ref{eq:sumavgdelay1}) into (\ref{eq:CAPmin}), we can calculate minimum required  CBAP duration for the completion of given number of requests. If we assume that on an average each STA should be able to complete on request in the allotted CBAP time, $N$ can be replaced by total number of STAs $n_k$ in that QO level to calculate the minimum average required CBAP time for that QO level.

%The average SF throughput is the average over beacon period, CAP and CTAP periods which can be given by,
%\begin{equation}
%S_{avg}=\frac{T_{Beacon}S_{Beacon}+T_{CAP}S_{CAP}+T_{CTAP}S_{CTAP}}{T_{Beacon}+T_{CAP}+T_{CTAP}}
%\end{equation}
\section{Numerical results and discussion}\label{discussion}
We assume 60\,GHz PCP/AP placed at the centre in the ceiling of room. We simulate a conference room environment (radius=10\,m) for different number of STAs as shown in Fig.~\ref{fig_system_model}. Beamwidth of all the STAs is assumed to be equal and taken as 60 degree. Location of an STA is identified by the Euclidean distance and angle from the PCP/AP. The Euclidean distances of STAs from the PCP/AP are uniformly distributed in the range [1,10\,m], while angles are generated from a Gaussian distribution with mean equal to 180 degree and a standard deviation of 90 degree. Thus, device distribution ensures uneven distribution of devices (some regions  are densely packed while some are not). All the parameters values are taken from the IEEE 802.11ad~\cite{IEEE802.11ad} and listed in Table~\ref{table_CAP}. For the RTS, DMG CTS and ACK frames, DMG control PHY (27.5\,Mbpas) is used. Data frames are transmitted using mandatory MCS4 (1.15\,Gbps). Both, the minimum beamwidth $\Omega_{min}$ and differential beamwidth $\Delta\Omega$ are assumed equal to 20 degree. To calculate the minimum required CBAP duration, total number of request per QO level are taken equal to number of STAs in that QO level. For fixed beamwidth, per QO level beamwidth is equal to 90 degree while for adaptive beamwidth, each QO level has a different beamwidth and decided using Algorithm~\ref{ALG}. 

Fig.~\ref{fig:CAPcompare} shows the comparison of channel utilization during CBAP period for fixed beamwidth and adaptive beamwidth QO patterns.  Fig.\ref{fig:CAPthroughputAdapt} shows the CBAP channel utilization for fixed and adaptive beamwidth QO levels. It is clear that the adaptive beamwidth approach performs better than the fixed beamwidth allocation. The reason behind this is that adaptive beamwidth approach always tries to accommodate STAs in such a manner that maximum channel time is utilized. If the number of STAs are increased, throughput of fixed beamwidth approach deteriorates, but adaptive beamwidth approach is able to maintain a fair throughput. This is because when the number of STAs are increased, fixed beamwidth QO levels suffer increased collisions due to more number of STAs, while adaptive beamwidth approach is able to reduce the beamwidth in heavily populated regions and thus restricting the number of STAs under the each QO level. This minimizes the collisions and maintains a steady throughput which is 20-30~$\%$ more than the fixed beamwidth sectors. 

Fig.~\ref{fig:TCAPminAdapt} shows the minimum required CBAP duration for successful transmission of all the generated requests. It shows that using adaptive beamwidth approach, CBAP duration is considerably reduced as compared to fixed beamwidth approach, e.g., when the total number of STAs are $50$, required minimum CBAP duration is about 40-50~$\%$ lesser for the adaptive beamwidth approach. It is evident from the results that as the number of STAs increase, the difference between the output of two approaches widens (for both channel utilization and minimum required CBAP duration). Hence it is clear that adaptive beamwidth QO levels help to reduce the required CBAP duration and maintain the steady channel utilization with increase in number of STAs. Therefore, in the regions having high STA density, it is better to narrow down the QO levels beamwidths to avoid the collisions. Hence, spatial distribution of nodes should be taken into account while granting channel access during CBAP through multiple QO levels of PCP/AP.
%
%
%\begin{figure*}[!t]
%% ensure that we have normalsize text
%\normalsize
%% Store the current equation number.
%\setcounter{MYtempeqncnt}{\value{equation}}
% Set the equation number to one less than the one
% desired for the first equation here.
% The value here will have to changed if equations
% are added or removed prior to the place these
% equations are referenced in the main text.
%\setcounter{equation}{18}
%\begin{equation}\label{eq22}
% b_{0,0}=\frac{2(1-2p)(1-p)}{W_0(1-p)(1-(2p)^{m+1})+(1-2p)(1-p^{m+1}+(2^mW_0+1)(1-p^{H-m}))}
%\end{equation}
%% Restore the current equation number.
%\setcounter{equation}{\value{MYtempeqncnt}}
%% IEEE uses as a separator
%\hrulefill
%% The spacer can be tweaked to stop underfull vboxes.
%%\vspace*{4pt}
%%
%\end{figure*}
\vspace*{-10pt}
\section{Conclusion}\label{remark} 
In this paper, we investigated the CBAP part of IEEE 802.11ad MAC protocol using directional antennas. Detailed performance analysis of time sharing CSMA/CA mechanism considering different QO levels is provided. It is shown that adaptive QO level beamwidth selection algorithm considering spatial distribution of STAs is highly effective and helped to achieve significant improvement in MAC throughput. It is shown that performance of fixed beamwidth approach deteriorates with increase in number of STAs while adaptive beamwidth approach is able to maintain a steady throughput. Numerical evaluations have shown that as compared to fixed beamwidth allocation, adaptive beamwidth allocation improves channel utilization up to 20-30$\%$ and reduces required CBAP duration by 40-50$\%$ for the given network conditions.

\bibliographystyle{IEEEtran}
\bibliography{Adaptreferences}

\begin{thebibliography}{10}
\providecommand{\url}[1]{#1}
\csname url@samestyle\endcsname
\providecommand{\newblock}{\relax}
\providecommand{\bibinfo}[2]{#2}
\providecommand{\BIBentrySTDinterwordspacing}{\spaceskip=0pt\relax}
\providecommand{\BIBentryALTinterwordstretchfactor}{4}
\providecommand{\BIBentryALTinterwordspacing}{\spaceskip=\fontdimen2\font plus
\BIBentryALTinterwordstretchfactor\fontdimen3\font minus
  \fontdimen4\font\relax}
\providecommand{\BIBforeignlanguage}[2]{{%
\expandafter\ifx\csname l@#1\endcsname\relax
\typeout{** WARNING: IEEEtran.bst: No hyphenation pattern has been}%
\typeout{** loaded for the language `#1'. Using the pattern for}%
\typeout{** the default language instead.}%
\else
\language=\csname l@#1\endcsname
\fi
#2}}
\providecommand{\BIBdecl}{\relax}
\BIBdecl

\bibitem{iee:smulders}
P.~Smulders, ``Exploiting the 60 ghz band for local wireless multimedia access:
  prospects and future directions,'' \emph{Communications Magazine, IEEE},
  vol.~40, no.~1, pp. 140--147, 2002.

\bibitem{iee:diffraction}
M.~Jacob, S.~Priebe, R.~Dickhoff, T.~Kleine-Ostmann, T.~Schrader, and
  T.~Kurner, ``Diffraction in mm and sub-mm wave indoor propagation channels,''
  \emph{Microwave Theory and Techniques, IEEE Transactions on}, vol.~60, no.~3,
  pp. 833--844, 2012.

\bibitem{iee:ref2}
S.~Singh, F.~Ziliotto, and U.~Madhow, ``Blockage and directivity in
  60\,\textsc{GH}z wireless personal area networks: From cross layer model to
  multi-hop mac design,'' \emph{IEEE J. Sel. Areas Commun.}, vol.~27, pp.
  1400--1413, Oct. 2009.

\bibitem{beamswitching_xeuli}
X.~An, C.-S. Sum, R.~Prasad, J.~Wang, Z.~Lan, J.~Wang, R.~Hekmat, H.~Harada,
  and I.~Niemegeers, ``Beam switching support to resolve link-blockage problem
  in 60 ghz wpans,'' in \emph{Personal, Indoor and Mobile Radio Communications,
  2009 IEEE 20th International Symposium on}, 2009, pp. 390--394.

\bibitem{diversity}
M.~Park and H.~K. Pan, ``A spatial diversity technique for ieee 802.11ad wlan
  in 60 ghz band,'' \emph{Communications Letters, IEEE}, vol.~16, no.~8, pp.
  1260--1262, 2012.

\bibitem{relaybyZulkuf}
Z.~Genc, G.~Olcer, E.~Onur, and I.~Niemegeers, ``Improving 60 ghz indoor
  connectivity with relaying,'' in \emph{ICC, IEEE}, 2010, pp. 1--6.

\bibitem{iee:ref7}
\BIBentryALTinterwordspacing
B.~L. Dang, M.~G. Larrode, R.~V. Prasad, I.~Niemegeers, and A.~M.~J. Koonen,
  ``Radio-over-fiber based architecture for seamless wireless indoor
  communication in the 60ghz band,'' \emph{Comput. Commun.}, vol.~30, no.~18,
  pp. 3598--3613, Dec. 2007. [Online]. Available:
  \url{http://dx.doi.org/10.1016/j.comcom.2007.08.041}
\BIBentrySTDinterwordspacing

\bibitem{iee:toward}
B.~L. Dang, R.~V. Prasad, I.~G. Niemegeers, M.~G. Larrode, and A.~M.~J. Koonen,
  ``Toward a seamless communication architecture for in-building networks at
  the 60 ghz band,'' in \emph{LCN'06}, 2006, pp. 300--307.

\bibitem{iee:IEEE802.15.3c}
``\textsc{IEEE} $802.15.3c$ working group, tgc3,,'' \emph{Report}.

\bibitem{IEEE802.11ad}
``{Draft standard- Part 11: Wireless LAN \textsc{MAC} and \textsc{PHY}
  specifications - Amendment 4: Enhancements for very high throughput in the
  60\,GHz band},'' \emph{IEEE P802.11adTM/D9.0}, July 2012.

\bibitem{harada}
C.~W. Pyo and H.~Harada, ``Throughput analysis and improvement of hybrid
  multiple access in ieee 802.15.3c mm-wave wpan,'' \emph{Sel Areas Comm., IEEE
  Journal on}, vol.~27, no.~8, pp. 1414--1424, 2009.

\bibitem{optbeam_leonardo}
L.~Goratti, T.~Wysocki, M.~Akhavan, J.~Lei, H.~Nakase, and S.~Kato, ``Optimal
  beamwidth for beacon and contention access periods in ieee 802.15.3c wpan,''
  in \emph{PIMRC, IEEE}, 2010, pp. 1395--1400.

\bibitem{11adcoperative}
Q.~Chen and et~al., ``{Directional Cooperative MAC Protocol Design and
  Performance Analysis for IEEE 802.11ad WLANs},'' \emph{Vehicular Technology,
  IEEE Trans. on}, vol.~62, no.~6, 2013.

\bibitem{iee:ref5}
G.~Bianchi, ``Performance analysis of the \textsc{IEEE} $802.11$ distributed
  coordination function,'' \emph{IEEE J. Sel. Areas Commun.}, vol.~18, pp.
  535--547, March 2000.

\bibitem{iee:ref8}
P.~Chatzimisios, A.~Boucouvalas, and V.~Vitsas, ``\textsc{IEEE} 802.11 packet
  delay-a finite retry limit analysis,'' in \emph{GLOBECOM '03. IEEE}, dec.
  2003.

\bibitem{iee:beamcodebook}
J.~Wang, Z.~Lan, C.~woo Pyo, T.~Baykas, C.-S. Sum, M.~Rahman, J.~Gao,
  R.~Funada, F.~Kojima, H.~Harada, and S.~Kato, ``Beam codebook based
  beamforming protocol for multi-gbps millimeter-wave wpan systems,''
  \emph{Sel. Areas Comm., IEEE Journal on}, vol.~27, no.~8, pp. 1390--1399,
  2009.

\bibitem{iee:MiuraOKU10}
A.~Miura, M.~Ohira, S.~Kitazawa, and M.~Ueba, ``60-ghz-band switched-beam
  eight-sector antenna with sp8t switch for 180 azimuth scan,'' \emph{IEICE
  Transactions}, vol. 93-B, no.~3, pp. 551--559, 2010.

\bibitem{beaconImplement}
S.~F.~A. Shah, S.~Srirangarajan, and A.~Tewfik, ``Implementation of a
  directional beacon-based position location algorithm in a signal processing
  framework,'' \emph{Wireless Comm., IEEE Trans. on}, vol.~9, no.~3, pp.
  1044--1053, 2010.

\end{thebibliography}
\end{document}